\documentstyle[11pt,newpasp,twoside,epsf]{article}
\def\edcomment#1{\iffalse\marginpar{\raggedright\sl#1\/}\else\relax\fi}
\reversemarginpar

\begin{document}
\title{Constraining the third dredge-up via carbon stars in the Magellanic
Clouds}
 \author{L\'eo Girardi}
\affil{Osservatorio Astronomico di Trieste, Italy}
\author{Paola Marigo}
\affil{Dipartimento di Astronomia, Universit\'a di Padova, Italy}

\begin{abstract}
We use the available data for Magellanic Cloud carbon stars to 
constrain the efficiency of the third dredge-up process in 
TP-AGB models.
We show that star counts in LMC clusters provide 
quite stringent limits to the lifetime of the C-star phase,
with a duration between $2$ and $3$ Myr for stars in
the mass range from $\sim1.5$ to $2.8\,M_\odot$. 
Together with the luminosity functions of field C stars, 
this information allows us to re-calibrate
the third dredge-up parameters $\log T_{\rm b}^{\rm dred}$ 
and $\lambda$ in TP-AGB models that include variable molecular
opacities (Marigo 2002). Preliminary results are presented here.

\end{abstract}

\section{Introduction}
During the thermally-pulsing (TP-) AGB phase of low- and intermediate-mass 
stars, several ``third dredge-up'' episodes --
followed by hot-bottom burning in the most massive stars -- 
cause the surface pollution with He and CNO elements of primary origin, 
which are eventually ejected into the interstellar medium. 
Although the occurrence of the third dredge-up is well 
established from theory, its efficiency is still uncertain. Also uncertain 
is the total duration of the TP-AGB phase, determined by both the dredge-up 
efficiency in reducing the core mass, and the mass-loss efficiency in 
peeling the star of its envelope. These uncertainties directly affect the 
estimates of the He and CNO chemical yields, as well as the predicted 
abundances of planetary nebulae. 

The most evident effect of the third dredge-up is the formation of carbon (C) 
stars when the surface C/O ratio increases above 1. Therefore,
C stars in resolved stellar populations constitute the best tool for 
calibrating the third dredge-up efficiency and hence constraining TP-AGB 
models and their chemical yields (see e.g. Marigo et al. 1999, and
Marigo 2001).

\begin{figure}  
\plottwo{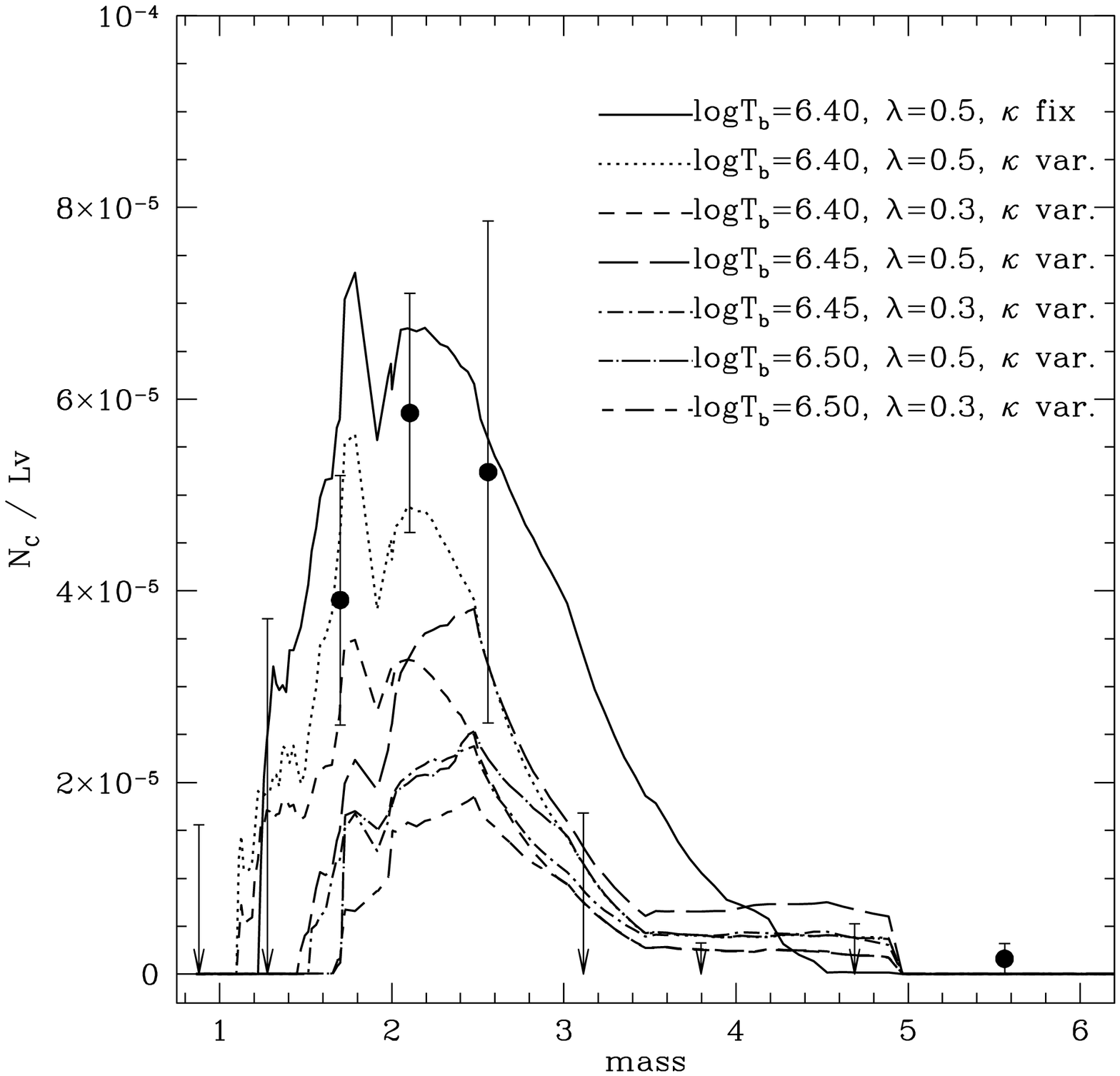}{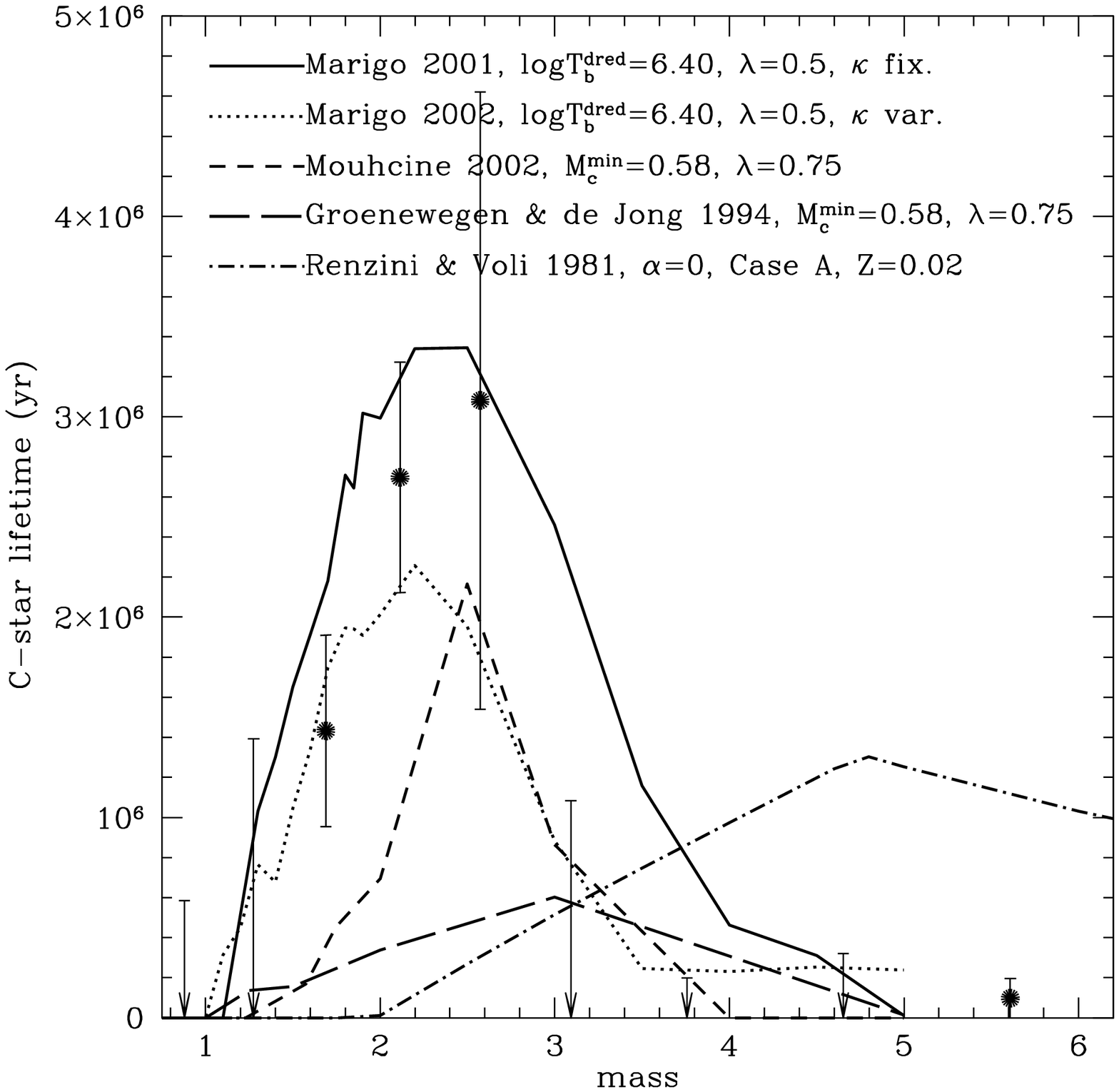}
\caption{
{\bf Panel a:} The number of C stars in LMC clusters normalised to 
their integrated $V$ luminosity, $N_{\rm C}/L_V$, 
as a function of turn-off mass. 
The data (full dots; see Frogel et al.\ 1990 and
Marigo et al.\ 1996) well constrain the mass range 
between 1.5 and 2.8 $M_\odot$, outside just 
upper limits (arrows) can be defined.
The several lines plot $N_{\rm C}/L_V$ 
for models with different opacities and dredge-up parameters. 
{\bf Panel b:} The same data translated in terms of C-star 
lifetimes, and compared to predictions by different TP-AGB models. 
}
\label{fig_1}
\end{figure}

Recently, Marigo (2002, and talk in this meeting) 
showed that the adoption of variable molecular opacities -- 
suitably coupled to the varying envelope 
composition along the TP-AGB -- dramatically changes the evolutionary 
properties of TP-AGB models, allowing a much better agreement with 
observed $T_{\rm eff}$ and C/O values for C stars. 
The improved models make it necessary a 
re-calibration of the third dredge-up efficiency.
With this aim, we are currently analysing 
the C star data for the Magellanic Clouds. 

\section{A few initial results}

One main evolutionary effect of 
variable molecular opacities is a reduction of $T_{\rm eff}$ 
values for C stars, which then causes an earlier onset of
the superwind regime and reduces their lifetimes. 

To check the lifetimes of the new TP-AGB models, we have used
the number of C stars in LMC star clusters normalised to their 
integrated $V$-band luminosity, $N_{\rm C}/L_V$ (Fig.\,1a). 
This plot gives an almost direct and model-independent
measure of the typical C-star lifetimes. 
The data (dots with error bars) put clear constraints for masses 
between 1.5 and 2.8 $M_\odot$, whereas at higher/lower masses just 
upper limits (arrows) can be defined.
Models with more efficient dredge-up (lower $\log T_{\rm b}^{\rm dred}$ 
and/or higher $\lambda$) are clearly favoured by the comparison
with these data. The same data indicate that C-star lifetimes vary 
between $2$ and $3$ Myr, for stars in the mass interval from 
$\sim1.5$ to $\sim2.8\,M_\odot$ (Fig.\,1b). It should be noticed
that several models in the literature present lifetimes 
significantly shorter than these values.
In models with variable molecular opacities, the observed lifetimes 
are attained provided that efficient dredge-up is favoured at 
low-masses -- for instance with 
the dredge-up parameters $\log T_{\rm b}^{\rm dred}=6.40$ 
and $\lambda=0.5$. These are the same values calibrated 
by Marigo et al. (1999). 

Models with these parameters tend to reproduce reasonably well 
the luminosities of individual C stars 
observed in LMC clusters, although such data is scarse and do 
not cover the complete age interval of interest. 
Instead, the LMC C-star luminosity function is slightly 
better reproduced by models with a later onset of the dredge-up, 
i.e.\ $\log T_{\rm b}^{\rm dred}=6.45$ (Fig.~2). 
For the SMC, the mean peak luminosity is well 
reproduced with $\log T_{\rm b}^{\rm dred}=6.40$. 
We are presently testing different combinations for 
$\log T_{\rm b}^{\rm dred}$ and $\lambda$, 
in order to find dredge-up parameters that best fit all
the available data.

\begin{figure}  
\plottwo{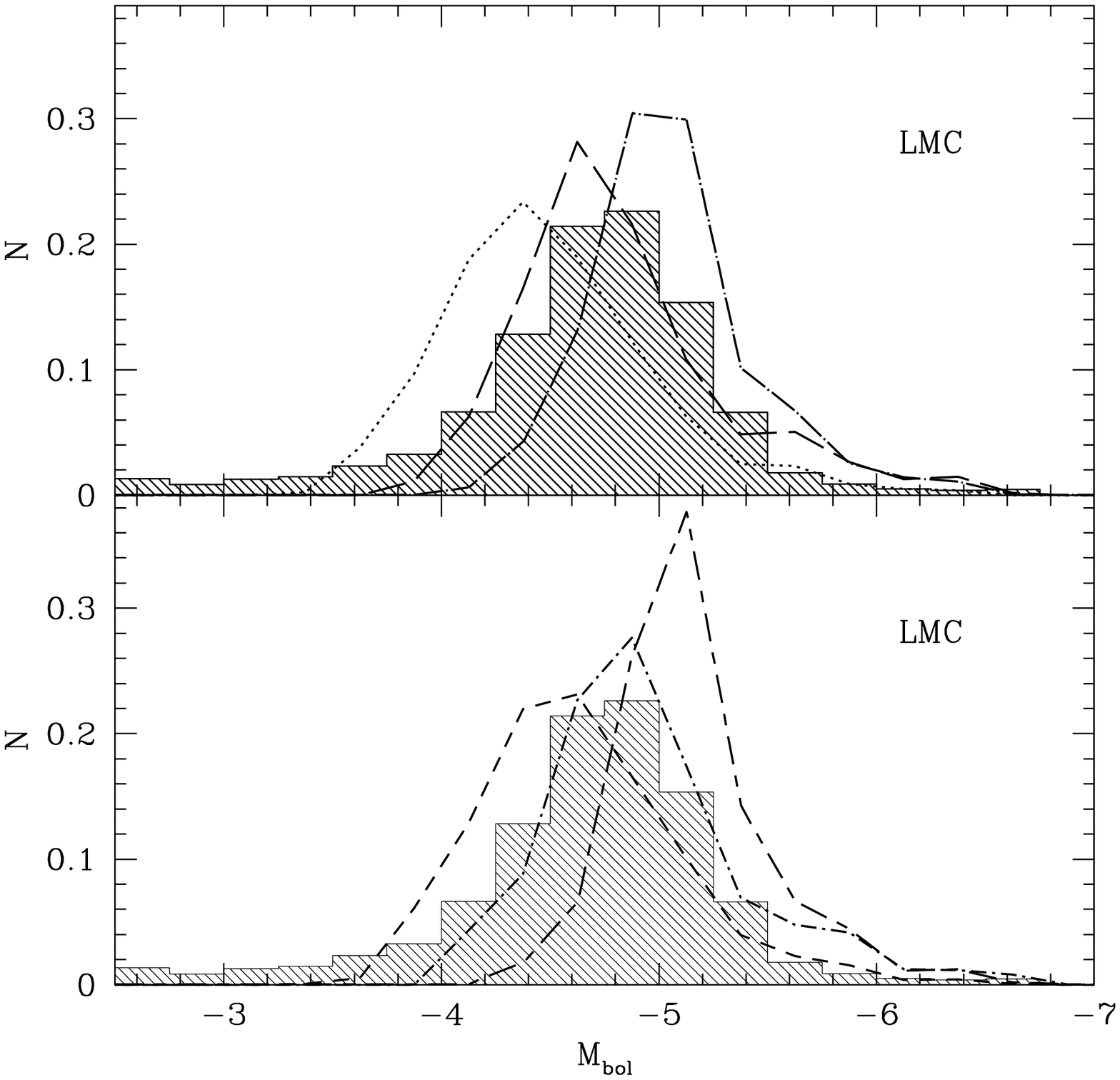}{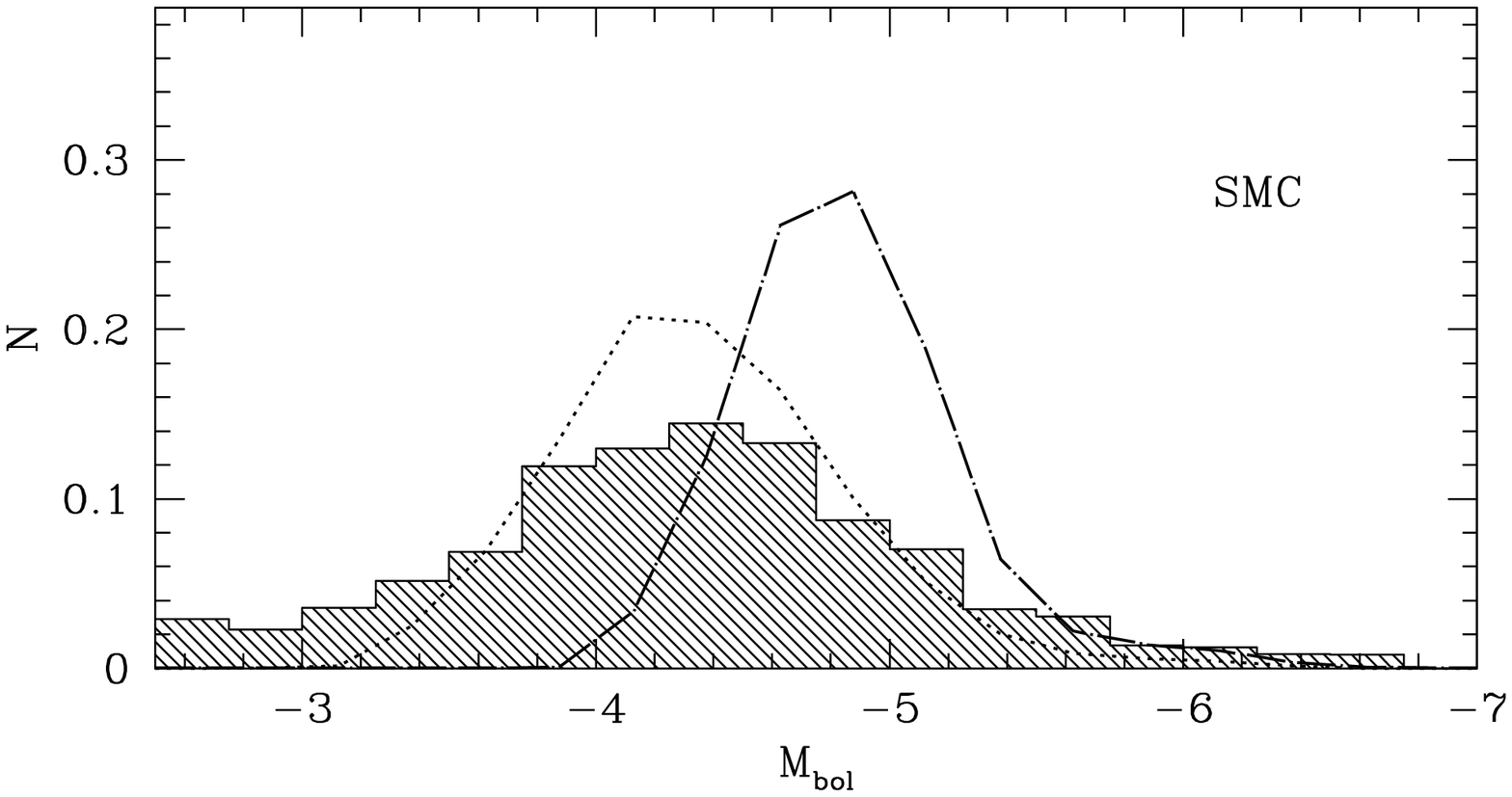}
\caption{
The observed C star luminosity functions (CSLF) in the 
Magellanic Clouds (Groenewegen 2002; shaded areas), as compared to the 
synthetic ones derived from models with different dredge-up parameters 
(lines, with the same meaning as in Fig.~1a). 
The results depend little on details like the assumed star 
formation rate and age-metallicity relation. Models with 
$\log T_{\rm b}^{\rm dred}=6.45$ better reproduce the CSLF for the LMC. 
}
\label{fig_2}
\end{figure}

Our preliminar calibrations are being checked in comparison with additional 
data for AGB stars in Local Group galaxies. Once the calibration is 
carried out, improved sets of chemical yields for TP-AGB models will be 
derived, as well as the related theoretical isochrones and synthetic CMDs
(Marigo et al.\ 2003).

\end{document}